\documentclass[aps, pra,twocolumn,groupedaddress,showpacs,nofootinbib,floatfix]{revtex4}
\usepackage{amsmath, amsthm, amssymb}
\usepackage{array}
\usepackage{amscd}
\usepackage{dsfont}
\usepackage{setspace}
\usepackage{graphicx}
\usepackage{color}
\usepackage[latin1]{inputenc}
\usepackage[english]{babel}
\usepackage{dcolumn}
\begin{document}

\title{Disentanglement and decoherence from classical non-Markovian noise:\\
Random telegraph noise}
\author{Dong Zhou, Alex Lang, and Robert Joynt}
\date{\today}
\affiliation{Physics Department, University of Wisconsin-Madison, Madison, Wisconsin
53706, USA}

\begin{abstract}
We calculate the two-qubit disentanglement due to classical random
telegraph noise using the quasi-Hamiltonian method. This allows us to obtain
analytical results even for strong coupling and mixed noise, important when
the qubits have tunable working point. We determine when entanglement sudden
death and revival occur as functions of qubit working point, noise coupling
strength and initial state entanglement. For extended Werner states, we show
that the concurrence is related to the difference of two functions: one is
related to dephasing and the other longitudinal relaxation. A\ physical
intepretation based on the generalized Bloch vector is given: revival only
occurs for strongly-coupled noise and comes from the angular motion of the
vector.   
\end{abstract}

\pacs{03.65.Yz,03.67.Mn,75.10.jm,85.25.Cp}
\maketitle

\section{Introduction}

Entanglement is a property that sets quantum systems apart from their
classical counterparts \cite{Horodecki_2009}. In recent years, it has drawn
great attention as an important resource for quantum information processing
and communication, such as quantum cryptography \cite{crypto}, dense coding 
\cite{Bennett_1992}, teleportation \cite{Bennett_1993} and exponential
speed-up of certain computational tasks \cite{algo}. Interactions with the
noisy environment inevitably degrade quantum coherence and thus
entanglement. It has been shown that although local (one-body) coherence
decays continuously, global coherence (entanglement) may terminate abruptly
in a finite time, a phenomenon known as entanglement sudden death (ESD) \cite
{Yu_2009_review}. To date ESD has been demonstrated in two different
experiments \cite{ESD_exp}.

In the theoretical investigations of ESD, the Markov approximation is
commonly used; namely, the environmental noise has short, or rather
instantaneous, self-correlations \cite{Yu_Eberly, Yu_2006,
Santos_2006,Ficek_2006}. Non-Markovian noise, however, is widely observed in
solid-state systems \cite{Kogan, Weissman_1988} and may even serve as the
dominant source of decoherence \cite{1/f, Nakamura_2002, Yoshihara_2006,
Kakuyanagi_2007, Khaetskii_2002}. It is thus crucial to extend the current
understanding of ESD to the presence of non-Markovian environments. It is
known that for noninteracting qubits coupled to their own independent environments, nonmonotonic time dependence of the
entanglement can only occur with non-Markovian noise \cite
{Bellomo_2007,Bellomo_2008, Dajka_2008}.
{ Mazzola \textit{et.al.} pointed out that a common structured non-Markovian reservoir protracts the disentanglement process and enriches the revivial \cite{Mazzola}. }
Yu \textit{et.al.} considered a
pure dephasing classical Ornstein-Uhlenbeck noise model and found the
short-time behavior of the entanglement evolution is markedly modified \cite
{Yu_2009}. Some progress has been made in modeling the behavior of
interacting 2-qubit systems in the presence of charge noise \cite{Hollenberg}.  
For the most part, however, the evolution of entanglement in a
non-Markovian environment is still an open question.

Pure dephasing noise models \cite{Yu_2002, Yu_2006, Roszak_2006,
Ann_2007,Yu_2009, Cao_2008, Dajka_2008} and variations of the
Jaynes-Cummings model \cite{Yu_Eberly,Bellomo_2007,Bellomo_2008,Mazzola} have been
the standard testbeds for ESD. Our work however, is motivated by the recent
discoveries in superconducting qubit designs \cite
{Nakamura_2002,Yoshihara_2006,Kakuyanagi_2007} where the working point of
the qubit is tunable: an arbitrary mixture of dephasing and relaxational
noise is possible.  In these architectures, it is found that $1/f$ noise is
the dominant source of decoherence. Thus a thorough understanding of those
experiments requires models that a) deal with non-Markovian noise,
especially random telegraph noise (RTN) \cite{Kogan} which are the basic
building blocks of $1/f$ noise; b) treat dephasing together with relaxation 
\cite{Galperin, Paladino, Joynt_2009, Zhou_2009,Cheng_2008}.

In this paper we use the quasi-Hamiltonian method \cite{Cheng_2008,
Joynt_2009, Zhou_2009} to investigate bipartite disentanglement of two
independent qubits caused by uncorrelated sources of classical RTN.  This
method is extremely powerful for these types of problems and we will be able
to obtain many analytic results, in an area of research dominated by
numerical studies.  Four issues are addressed. Firstly, it is known that
RTNs can be put into two categories according to the ratio of their
switching rates and coupling strengths to the qubit, namely the
weakly-coupled (fast, Markovian) ones and strongly-coupled (slow, non-Markovian) ones \cite
{Galperin,Joynt_2009, Zhou_2009}. We thus seek for qualitative differences
in the disentanglement caused by these two types of RTNs. Secondly, we
exploit the working point of the qubit to see if this extra degree of
freedom affects ESD and revival. This is particularly valuable when the
working point can be varied, since the number of strongly- and
weakly-coupled noise sources can actually be tuned.  Thirdly, some
entangled states are known to be more robust against disentaglement than
others \cite{Yu_2002}. We thus examine different initial states, both pure
(generalized Bell states) and mixed (extended Werner states). Finally, we
compare two noise models, the two-one model where only one qubit is subject
to a RTN source and the two-two model where both qubits are attached to RTNs
individually. This allows us to see the effect of noise locality on
entanglement, a global property.

There are two distinct physical effects that lead to disentanglement.  One
is the movement of entangled states toward product states: this can happen
even in the absence of noise.  Second is the movement of entangled states
towards completely mixed states: this requires noise.  We will separate
these two routes to disentanglement as far as we can by considering pure
entangled initial states and mixed entangled initial states;  in the latter
case, the distance from purity can be parametrized if the mixed states are
chosen as generalized Werner states. 

We also propose the magnitude of the generalized Bloch vector $\left\vert 
\vec{n}\right\vert $ as an appropriate purity measure for multi-qubit
states.  In the single qubit case, states on the surface of the Bloch
sphere ($\left\vert \vec{n}\right\vert =1$) are pure, while states at the
origin ($\left\vert \vec{n}\right\vert =0$) are completely mixed.  In the
2-qubit case, the situation is not quite so simple, since the set of
admissible states is not spherical.  However, as we shall show, the pure
states lie on the surface $\left\vert \vec{n}\right\vert =\sqrt{3}$. In the
2-qubit state, we can then separate the sources of entanglement by computing
both the concurrence and $\left\vert \vec{n}\right\vert $.

The paper is organized as follows. In Sec. \ref{sec:model}, we introduce the
model Hamiltonian and apply the quasi-Hamiltonian method to reduce the
two-qubit problem to the single-qubit problem. In Sec. \ref{sec:functional}
we discuss the dephasing and relaxation behavior of single qubit as function
of the qubit working point and RTN property. In Sec. \ref{sec:conc} we
define the two-qubit entanglement measure. In Sec. \ref{sec:2_1_model} and 
\ref{sec:2_2_model}, we solve the two-one and two-two models. In Sec. \ref
{sec:phase}, we discuss the geometrical interpretation of the Bloch vector
and qualitative difference between weak and strong coupling, and give
overall conclusions.

\section{model}

\label{sec:model} The Hamiltonian of the system is given by 
\begin{align}
H(t) = -\frac{1}{2}\sum_{K=A,B} \left[B_0\sigma^K_{z}+s^K(t) \vec
g_K\cdot\vec\sigma^K \right].
\end{align}
where $A,B$ refers to the two qubits, $B_0$ is the energy splitting of the
qubits between the ground and excited states, $\vec g$ is the coupling of
the RTN to the qubit, and $s(t)$ is the RTN sequence that switches between
the values $\pm1$ with an average switching rate $\gamma$. Here $
\vec\sigma=[\sigma_1;\sigma_2;\sigma_3]$ is the triad of the Pauli matrices.

This Hamiltonian is general enough to describe any qubit subject to
classical RTNs. For a superconducting flux qubit, $B_{0}=\sqrt{\varepsilon
^{2}+\Delta ^{2}}$, where $\varepsilon $ is proportional to the applied flux
through the superconducting loop and $\Delta $ is the Josephson coupling,
which are the energy difference and tunneling splitting between the two
physical states. In this case $\theta =\tan^{-1}(\Delta /\varepsilon )$
is independent of $K$ and the angle $\theta$
 is called the working point of the qubit. Since
flux noise is along the $\varepsilon $ direction, $\theta $ is the angle
between the noise direction and the qubit eigenstate direction, and it
can be varied by changing the applied flux. For spin qubits, $\theta _{K}$
is simply the angle between the applied field and the magnetic noise field
of the $K$-th fluctuator.  For general problems, an arbitrary power and
angular $\left( \theta \right) $ spectrum can be obtained by superposing
noise sources.

In this paper we treat unbiased noise, so $\overline{s(t)}=0,$ where the
overbar denotes averaging over the noise distribution.  We also have 
\begin{equation}
\overline{s(t)s(t^{\prime})}=\exp \left( -2\gamma \left\vert t-t^{\prime
}\right\vert \right) .
\end{equation}
The noise autocorrelation function dies off exponentially, corresponding to
a Lorentzian power spectrum $S_\text{RTN}(\omega)=4\gamma g^2/(\omega^2+4\gamma^2)$. There are thus two time scales set by the RTN's
characteristics: the correlation time of the environment $\tau _{e}\sim
1/\gamma $ and the noise induced looping time of the qubit $\tau _{l}\sim
1/\left( g\cos \theta \right) $, where $\theta $ is the angle between the
energy axis $\hat{z}$ and the noise coupling direction $\hat{g}$ \cite
{Zhou_2009}. The relative lengths of these two
time scales are critical for the qubit decoherence and disentanglement.
Indeed, $\tau_e<\tau_\ell$ is effectively the weak-coupling (Markovian) 
region while $\tau_e>\tau_\ell$ is the strong-coupling (non-Markovian) region.

The density matrix $\rho _{AB}(t)$ is $4\times 4$ and can be expanded in the
generators $\mu _{i}$ of $SU(4)$ 
\begin{equation}
\rho _{AB}(t)=\frac{1}{4}\left( I_{4}+\sum_{i=1}^{15}n_{i}(t)~\mu
_{i}\right) ,  \label{eq:density_matrix}
\end{equation}
where $I_{4}$ is the $4\times 4$ unit matrix. $n_{i}(t)=\text{Tr~}[\rho
_{AB}(t)\mu _{i}]$ are the components of the generalized Bloch vector $\vec{n
}$; they are all real. The choice of the set of $15$ Hermitian matrices $
\left\{ \mu _{i}\right\} $ is a choice of basis.  We will take them to be 
\begin{equation}
\{\sigma _{a}\otimes \sigma _{b}\}~\backslash ~\{\sigma _{0}\otimes \sigma
_{0}\},\quad a,b=\{0,1,2,3\}
\end{equation}
where $\sigma _{0}$ is the $2\times 2$ identity matrix. We adopt the base-4
ordering convention such that $\mu _{1}=\sigma _{0}\otimes \sigma _{1}$, $
\mu _{2}=\sigma _{0}\otimes \sigma _{2}$, etc.  The $\mu _{i}$ are an
orthonormal basis for the density matrix space with the inner product $
\left\langle \mu ,\mu ^{\prime }\right\rangle =$Tr$\left[ \mu \mu ^{\prime
}\right] /4$. Note also that Tr $\mu _{i}=0$.

The $n_{i}\left( t\right) $ are not the most common way to characterize a
quantum state.  However, they have a direct physical meaning.  For
example, since $\mu _{10}=\sigma _{2}\otimes \sigma _{2}$, we have that 
\begin{align*}
n_{10}(t) = \text{Tr~} \left[\mu_{10}\rho_{AB}(t)\right] = \left\langle\sigma _{2}\otimes \sigma _{2}\right\rangle.
\end{align*}
Thus $n_{10}$ is the value of a certain spin-spin correlation function.

Furthermore, the $n_{i}\left( t\right) $ collectively form a measure of the
purity of the state \cite{Byrd}.  A pure state satisfies $\rho ^{2}=\rho$.
 Therefore any pure state satisfies $0=$Tr $\left( \rho -\rho ^{2}\right) =
\frac{3}{4}-\frac{1}{4}\left\vert \vec{n}\right\vert ^{2},$so $\left\vert \vec{n}
\right\vert =\sqrt{3}$.  At the other limit, the completely mixed state $
\rho =I_{4}/4$ gives $\vec{n}=0$ and Tr$\left( \rho -\rho ^{2}\right) =3/4$.
 However, not all states with $\left\vert \vec{n}\right\vert \leq \sqrt{3}$
respect positivity \cite{Byrd}.

Using the quasi-Hamiltonian method \cite{Cheng_2008,Joynt_2009}, the time
evolution of the quantum system in the presence of classical noise can be
cast into a time-dependent transfer matrix $T(t)$ acting on the generalized
Bloch vector, 
\begin{equation*}
\vec{n}(t)=T(t)~\vec{n}(0)
\end{equation*}
Note $T(t)$ is real but is not orthogonal once the average over noise
histories has been performed.  Thus nonorthogonality is a direct
consequence of the incoherent environment.

To simplify notation, we include $\mu _{0}=I_{4}$ in the generalized Bloch
vector, i.e. we define the 'extended' generalized Bloch vector as $
\underline{\vec{n}}=[n_{0};\vec{n}]$. Note $n_{0}(t)=1$ for all time. We
thus have 
\begin{equation*}
\underline{\vec{n}}(t)=\underline{T}(t)~\underline{\vec{n}}(0)
\end{equation*}
and the time evolution $\underline{T}(t)$ can be succinctly written as 
\begin{equation*}
\underline{T}(t)=\underline{R}^{A}(t)\otimes \underline{R}^{B}(t),
\end{equation*}%
where $\underline{R}^{K}(t)$, $K=A,B$ are the $4\times 4$ 'extended' single
qubit transfer matrices. They are derived from the single qubit transfer
matrices $R^{K}$ by padding the matrix: the $(00)$ entry is set to $1$
and the $(0i)$, $(i0)$ entries for $i=1,2,3$ are set to 0.  Now 
\begin{equation*}
R^{K}(t)=\left\langle x_{f}\right\vert \exp (-iH_{q}^{K}t)\left\vert
i_{f}\right\rangle ,\qquad K=A,B
\end{equation*}
and 
\begin{equation*}
H_{q}^{K}=-i\gamma +i\gamma \tau _{1}+\left[ B_{0}L_{z}+\tau _{3}\vec{g_{K}}
\cdot \vec{L}\right] .
\end{equation*}
Here $\left\vert i_{f}\right\rangle $ and $\left\vert x_{f}\right\rangle $
are related to the initial distributions of the RTN. We only consider
unbiased RTN in this paper: $\left\vert i_{f}\right\rangle =\left\vert
x_{f}\right\rangle =[1;1]/\sqrt{2}$.

If the qubit is not subject to any noise, the 'extended' single qubit
transfer matrix can be written as 
\begin{equation*}
\underline{R}_{0}(t)=
\begin{bmatrix}
1 & 0 & 0 & 0 \\ 
0 & \cos B_{0}t & \sin B_{0}t & 0 \\ 
0 & -\sin B_{0}t & \cos B_{0}t & 0 \\ 
0 & 0 & 0 & 1
\end{bmatrix} .
\end{equation*}
$\underline{R}_{0}(t)$ is orthogonal and thus conserves the length of the
Bloch vector. The time evolution of the single qubit Bloch vector is simply
a precession along the $\hat{z}$ direction with Larmor frequency $B_{0}$.

If the qubit is subject to a RTN, the transfer matrix can be calculated by
known methods \cite{Joynt_2009,Zhou_2009} and we have 
\begin{align}\label{eq:R_RTN}
\underline{R}_{\text{RTN}}(t)=
\begin{bmatrix}
1 & 0 & 0 & 0 \\ 
0 & \zeta \left( t\right) \cos B_{0}t & \zeta \left( t\right) \sin B_{0}t & 0
\\ 
0 & -\zeta \left( t\right) \sin B_{0}t & \zeta \left( t\right) \cos B_{0}t & 
0 \\ 
0 & 0 & 0 & e^{-\Gamma _{1}t}
\end{bmatrix} .
\end{align}
where $\zeta (t)$ characterizes the dephasing behavior of the qubit and $
\Gamma _{1}$ is the longitudinal relaxation rate. They will be discussed in
the next section. $\underline{R}_{\text{RTN}}$ is generally a nonorthogonal
matrix since both $\zeta \left( t\right) $ and $\exp (-\Gamma _{1}t)$
decrease with time.

As been pointed out by Bellomo \textit{et.al.} \cite{Bellomo_2007}, the
dynamics of $N$-qubit density matrix elements follows from the dynamics of
each single qubit, and it is essentially independent of the initial
condition of the total system. This feature is especially clear in our
formalism, since $\underline{T}(t)$ is constructed directly from tensor
products of single qubit transfer matrices $\underline{R}^{K}(t)$ and is
independent of the initial conditions.
{It is know that collective channels (common reservoir) can also lead to entanglement instability, either in the Markovian \cite{Ficek_2006} or non-Markovian environment \cite{Mazzola}. In this case, the two-qubit transfer matrix $\underline{T}(t)$ is no longer of product form of single-qubit transfer matrices.
The noise correlations glue up the single qubit Hilbert spaces and indirectly couple the two qubits \cite{Braun}.
}

The results in this paper have been calculated using the quasi-Hamiltonian method and have also been verified through numerical simulations. A single numerical "run" is calculated by generating a sequence of RTN and then exactly solving the density matrix for that given sequence. The final numerical simulation result is found by producing thousands of runs (40,000 for the figures in this paper) each with a different sequence of RTN, and then finding the average density matrix over all the runs. This allows us to numerical simulate the quasi-Hamiltonian results which are inherently averaged over all RTN sequences. The numerical and quasi-Hamiltonian simulations are in agreement to within round-off error and could be plotted on the same graph without any visible difference.

\section{single-qubit dephasing and relaxation}

\label{sec:functional} It is known that the fast (weakly-coupled, Markovian) RTNs and
slow (strongly-coupled, non-Markovian) RTNs have qualitatively different effects on the
single qubit time evolutions \cite{Joynt_2009, Zhou_2009, Galperin, Paladino}
.

The two types of RTNs are separated by the criterion $g\cos \theta =\gamma $
. For the fast ones ($\gamma >g\cos \theta $), we have $\tau _{l}>\tau _{e}$
and the memory of the environment is short comparing to the looping time.
Redfield theory \cite{Slichter_1996} applies in this case and both dephasing
and relaxation of the elements of the density matrix is exponential at long
times while the very short time behavior is quadratic \cite{Joynt_2009}.

For the non-Markovian RTN sources ($\gamma <g\cos \theta $), $\tau _{l}<\tau _{e}$,
the correlation of noise is long enough to make possible the looping of the
Bloch vector on the Bloch sphere. This looping manifests itself as
oscillations in the Free-Induction signal (FID) and Spin-Echo (SE) signals 
\cite{Zhou_2009}.

The function $\zeta \left( t\right) $ introduced in Eq.\ref{eq:R_RTN} is
directly related to the FID signal, i.e. $n_{\text{FID}}(t)=\cos
B_{0}t~\zeta (t)$. Physically, $\zeta \left( t\right) $ is the probability
for the single-qubit Bloch vector to return to its starting point on the
Bloch sphere in the rotating frame when no pulses are applied. \ We thus
call $\zeta \left( t\right) $ the dephasing function. Note $\zeta (t=0)=1$
and $\zeta (t\rightarrow \infty )=0$ if dephasing occurs.

For the Markovian RTN we have the well-known results: 
\begin{align}
\zeta (t)=e^{-\Gamma _{2}t},\label{eq:zeta_aw2}
\end{align}
where 
\begin{align}
\Gamma _{2}=& \frac{\Gamma _{1}}{2}+\frac{g^{2}\cos ^{2}\theta }{2\gamma },
\label{eq:Gamma_phi} \\
\Gamma _{1}=& \frac{2\gamma g^{2}\sin ^{2}\theta }{4\gamma ^{2}+B_{0}^{2}}.
\label{eq:G1_fast}
\end{align}

For the non-Markovian RTN 
\begin{align}
\zeta (t)=e^{-\gamma t}\left[ \cos (g\cos \theta t)+\epsilon _{1}\sin (g\cos
\theta t)\right]\label{eq:zeta_aw1}
\end{align}
and the longitudinal relaxation rate is 
\begin{align}
\Gamma _{1}=2\gamma \epsilon _{2}^{2}\sin ^{2}\theta ,\label{eq:G1_slow}
\end{align}
where $\epsilon _{1}=\gamma /g\cos \theta $ and $\epsilon _{2}=g/B_{0}$. 
In each case the dephasing function $\zeta \left( 0\right) =1$ and $\zeta
\left( t\rightarrow \infty \right) =0$.

Unlike the monotonic decay in the Markovian RTN case, $\zeta \left( t\right) $ is
oscillatory in the presence of non-Markovian RTN. It hits zero at discrete points in
time 
\begin{align}
\zeta \left( t_{\ell }\right) =0\text{ at }t_{\ell }=\frac{\pi \ell -\tan
^{-1}(1/\epsilon _{1})}{g\cos \theta },\qquad \ell =1,2,\ldots \label{eq:zeros_p}
\end{align}

Note in both strong and weak coupling region, the longitudinal qubit
relaxation can always be well characterized by a single exponential
coefficient $\Gamma _{1}$.

Finally, we note that there is an exact result for $\zeta \left( t\right) $
at the pure dephasing point $\theta =0$ \cite{Joynt_2009} 
\begin{align}
\zeta \left( t\right) =e^{-\gamma t}\left[ \cosh \left( \sqrt{\gamma
^{2}-g^{2}}t\right) +\frac{\sinh \left( \sqrt{\gamma ^{2}-g^{2}}t\right) }{
\sqrt{1-(\frac{g}{\gamma })^{2}}}\right] .\label{eq:zeta_pd1}
\end{align}

For Eq.\ref{eq:zeta_pd1}, if $g>\gamma $ (strong coupling), the hyperbolic
functions need to be replaced by the corresponding trigonometric functions 
\begin{align}
\zeta \left( t\right) =e^{-\gamma t}\left[ \cos (\sqrt{g^{2}-\gamma ^{2}}t)+
\frac{\sin (\sqrt{g^{2}-\gamma ^{2}}t)}{\sqrt{\left( \frac{g}{\gamma }
\right) ^{2}-1}}\right] .\label{eq:zeta_pd2}
\end{align}

Using the exact result, the zeros of $\zeta \left( t\right) $ are given by 
\begin{align}
t_{\ell }=\frac{\pi \ell -\tan ^{-1}\left( \sqrt{\left( \frac{g}{\gamma }
\right) ^{2}-1}\right) }{\sqrt{g^{2}-\gamma ^{2}}},\quad \ell =1,2,3,\ldots 
\end{align}
We see Eq.\ref{eq:zeros_p} is indeed the the correct behavior as $\gamma
\rightarrow 0$.

\section{concurrence}

\label{sec:conc} For bipartite entanglement, all entanglement measures are
compatible and we use concurrence \cite{Wootters_1998} for its ease of
calculation. The concurrence varies from $0$ for the disentangled state to $%
1 $ for the maximally entangled state. It is defined as $C^{AB}(t)=\max%
\{0,q(t)\}$, and 
\begin{align}
q(t) = {\lambda_1}-{\lambda_2}-{\lambda_3}-{\lambda_4}~,
\end{align}
where $\lambda_i$ are the square roots of the eigenvalues of the matrix $%
\rho_{AB}\tilde\rho_{AB}$ arranged in decreasing order and 
\begin{align}
\tilde\rho_{AB}=(\sigma^A_y\otimes\sigma_y^B)\rho_{AB}^*(\sigma^A_y\otimes%
\sigma^B_y),
\end{align}
where $\rho_{AB}^*$ is the complex conjugate of the density matrix $%
\rho_{AB}(t)$.

The product $\rho _{AB}\tilde{\rho}_{AB}$ can be expanded as 
\begin{equation*}
\rho _{AB}\tilde{\rho}_{AB}=\sum_{i,j=0}^{15}\frac{n_{i}n_{j}}{16}~\mu _{i}~
\tilde{\mu}_{j},
\end{equation*}
where $\tilde{\mu}_{j}=(\sigma _{y}\otimes \sigma _{y})\mu _{j}^{\ast
}(\sigma _{y}\otimes \sigma _{y})$. Note $\tilde{\mu}_{i}=-\mu _{i}$, if $
i=1,2,3,4,8,12$ and $\tilde{\mu}_{i}=\mu _{i}$ for other $i$'s.

Thus our formalism allows us to investigate the time evolution of bipartite
entanglement with the knowledge of the Bloch vector $n(t)$.

ESD occurs when $q(t)<0$ since the '$\max$' operation
forces $C^{AB}=0$ and $C^{AB}$ is not an analytic function. For the
situations considered in this paper, we find that $\lambda _{1}$ and $%
\lambda _{2}$ have the same long time limit and $\lambda _{3}=\lambda _{4}$.
Thus ESD happens if $\lim_{t\rightarrow \infty }\lambda _{3,4}(t)\neq 0$. In
this work, we also find $q(t)$ has a structure of $|\zeta |-\xi $, where $%
\xi $ is generally related to the initial state as well as the longitudinal
relaxation process. Thus ESD happens whenever $\xi (t\rightarrow \infty
)\neq 0$ since $\zeta (t)$ always decays to zero in the presence of
dephasing.

We also compute $|\vec n|(t)$, the magnitude of the generalized Bloch vector. This is a measure of purity in the single-qubit case. 
In the two-qubit case it appears to track $C^{AB}$ to a large extent, and might serve as a potential measure of both entanglement and purity for multiple qubit $(>2)$ systems when concurrence is no longer defined.

\section{two-one model}

\label{sec:2_1_model} In this section, we consider the case where only one
of the two qubits is subject to RTN. The 'extended' transfer matrix is thus
given by 
\begin{align}
\underline{T}(t) = \underline{R}_{\text{RTN}}^A(t)\otimes \underline{R}_0^B(t),
\end{align}

In this model qubit $B$ enjoys coherent time evolution and is stationary in
the rotating frame. Thus all disentanglement and decoherence of the
two-qubit system come from qubit $A$'s interaction with RTN.  We can think
of the qubit $B$ as forming a kind of reference frame for entanglement with
qubit $A$.

\subsection{Pure States}

We consider the generalized Bell states as initial states 
\begin{align}
\left|\Phi\right> =& \alpha\left|00\right>+\beta\left|11\right> \\
\left|\Psi\right> =& \alpha\left|01\right>+\beta\left|10\right>
\end{align}
where $\alpha$ is real positive, $\beta = \|\beta\|e^{i\delta}$ and $%
\alpha^2+\|\beta\|^2=1$. Note $\alpha=1/\sqrt{2}$, $\beta=\pm1/\sqrt 2$
gives the Bell bases $\left|\Phi^\pm\right>$ and $\left|\Psi^\pm\right>$.

With $\left\vert \Phi \right\rangle $ or $\left\vert \Psi \right\rangle $ as
initial state, a complete analytic solution is possible. 

The time evolution of the generalized Bloch vector for initial state $%
\left\vert \Phi \right\rangle $ is given by 
\begin{align}
n_{3}(t)=& 2\alpha ^{2}-1,\label{eq:pd_21} \\
n_{5}(t)=& 2\alpha ~\zeta  ~\left[ \cos (2B_{0}t)\text{Re}%
~\beta +\sin (2B_{0}t)\text{Im}~\beta \right] , \\
n_{6}(t)=& 2\alpha ~\zeta ~ \left[ \cos (2B_{0}t)\text{Im}%
~\beta -\sin (2B_{0}t)~\text{Re}~\beta \right] , \\
n_{9}(t)=& n_{6}(t), \\
n_{10}(t)=& -n_{5}(t), \\
n_{12}(t)=& n_{3}(t)e^{-\Gamma _{1}t}, \\
n_{15}(t)=& e^{-\Gamma _{1}t},
\end{align}
while the other $n_{i}(t)=0$. Here $\zeta \left( t\right) $ and $\Gamma _{1}$
are given by Eq.\ref{eq:zeta_aw2}, \ref{eq:G1_fast} or \ref{eq:zeta_aw1}, %
\ref{eq:G1_slow}, depending on the coupling of the RTN.  

With $\left\vert \Psi \right\rangle $ as initial state, the non-zero
components are 
\begin{align}
n_{3}(t)=& 1-2\alpha ^{2}, \\
n_{5}(t)=& 2\alpha ~\text{Re}~\beta~\zeta \left( t\right) ,\\
n_{6}(t)=& -2\alpha ~ \text{Im}~\beta~\zeta \left( t\right), \\
n_{9}(t)=& -n_{6}(t), \\
n_{10}(t)=& n_{5}(t), \\
n_{12}(t)=& -n_{3}(t)e^{-\Gamma _{1}t}, \\
n_{15}(t)=& -e^{-\Gamma _{1}t}.
\end{align}
{Note in this case the generalized Bloch is not dependent on $B_0$. This is because $\sigma_z^A+\sigma_z^B$ annihilates $\left|\Psi\right>$. }

\begin{figure}[tb]
\centering\includegraphics*[width=\linewidth]{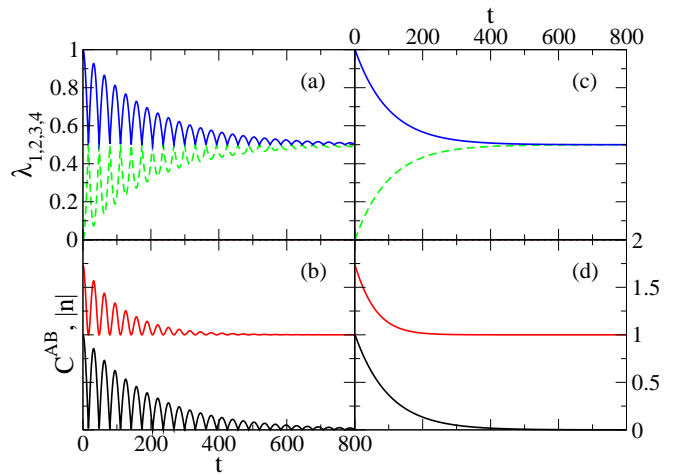}
\caption{Pure dephasing noise for the two-one model.  Top panels (a) and
(c): Square roots of eigenvalues of $\protect\rho \tilde{\protect\rho}(t)$.
 Bottom panels (b) and (d): magnitude of Bloch vector $n(t)$ above and
concurrence $C^{AB}(t)$ below.  These are calculated from Eqs. \ref{eq:pd_21}-\ref{eq:C_pure} and
confirmed by numerical simulations. Both qubits are operated at the \textbf{%
pure dephasing point} and only one of them is connected to the RTN, i.e. $%
g_{1}=0.1$, $g_{2}=0$, $\protect\theta =0$. The initial state is set to $%
\left\vert \Phi ^{+}\right\rangle $. In (a) and (b), $\protect\gamma =0.005$
and the RTN is in the strong-coupling region while in (c) and (d), $\protect%
\gamma =0.5$ and the RTN is in the weak-coupling region. $\protect\lambda %
_{1}$ is plotted as a solid blue line, $\protect\lambda _{2}$ as a dashed
green line; $\protect\lambda _{3}=\protect\lambda _{4}=0$. $n(t)$ is plotted
as a solid red line and $C^{AB}(t)$ as a solid black line.  Time is in the
unit $1/B_{0}$.  These results are exact.}
\label{fig:pd_Bell}
\end{figure}

For both initial states, the square roots of the eigenvalues of $\rho _{AB}%
\tilde{\rho}_{AB}$ are 
\begin{align}
\lambda _{1}=& \frac{\alpha \Vert \beta \Vert }{2}\left( 1+e^{-\Gamma
_{1}t}+2|\zeta |\right) , \\
\lambda _{2}=& \frac{\alpha \Vert \beta \Vert }{2}\left( 1+e^{-\Gamma
_{1}t}-2|\zeta |\right) , \\
\lambda _{3}=& \lambda _{4}=\frac{\alpha \Vert \beta \Vert }{2}(1-e^{-\Gamma
_{1}t}).
\end{align}%
which gives the concurrence as 
\begin{align}
C^{AB} &=2\alpha \sqrt{1-\alpha ^{2}}\max \left\{ 0,~~|\zeta(t)|-\frac{
1-e^{-\Gamma _{1}t}}{2}\right\}\notag \\
&=2\alpha \sqrt{1-\alpha ^{2}}\max \left\{ 0,~~|\zeta(t)|-\xi \left( t\right)
\right\} .\label{eq:C_pure}
\end{align}
The concurrence has a remarkable form.  $\xi \left( t\right) \geq 0$ is
only related to the longitudinal relaxation rate $\Gamma _{1}$ and varies
between $0$ for the unrelaxed state and $1/2$ for the fully relaxed state;
it describes relaxation only, while $\zeta \left( t\right) $ is related only
to dephasing.  Entanglement exists only when $\left\vert \zeta \right\vert
>\xi$.  The effects of dephasing (the decrease of $\left\vert \zeta
\right\vert $) and relaxation (the increase of $\xi )$ are additive, and
they race to disentangle the state.  However, $\zeta \left( t\right) $ can
be oscillatory, and then revival of entanglement is possible.  Once the
envelope of $\zeta \left( t\right) $ is less than $\xi \left( t\right)$,
entanglement is gone for good. 

As a function of working point $\theta $, $\Gamma _{1}$ is finite except for 
$\theta =0$: \textit{the pure dephasing point}.  At this point $\Gamma
_{1}=0$ and $\xi \left( t\right) =0$. Two of the four eigenvalues vanish: $
\lambda _{3,4}=0$.  If $\zeta \left( t\right) $ is a monotonic function
(weak coupling), then $C^{AB}=2\alpha \sqrt{1-\alpha ^{2}}|\zeta (\theta
=0)|>0$ at all finite times and ESD does not occur. This is seen in Fig. %
\ref{fig:pd_Bell} (c,d).  If $\zeta \left( t\right) $ is oscillatory
(strong coupling) then ESD and revival occurs, as seen in Fig. \ref%
{fig:pd_Bell} (a,b).  In fact the revival happens an infinite number of
times, since the envelope is exponential.  We again note that the result at
the pure dephasing point is exact; see Eqs. \ref{eq:zeta_pd1} and \ref{eq:zeta_pd2}.

\begin{figure}[tb]
\includegraphics*[width=\linewidth]{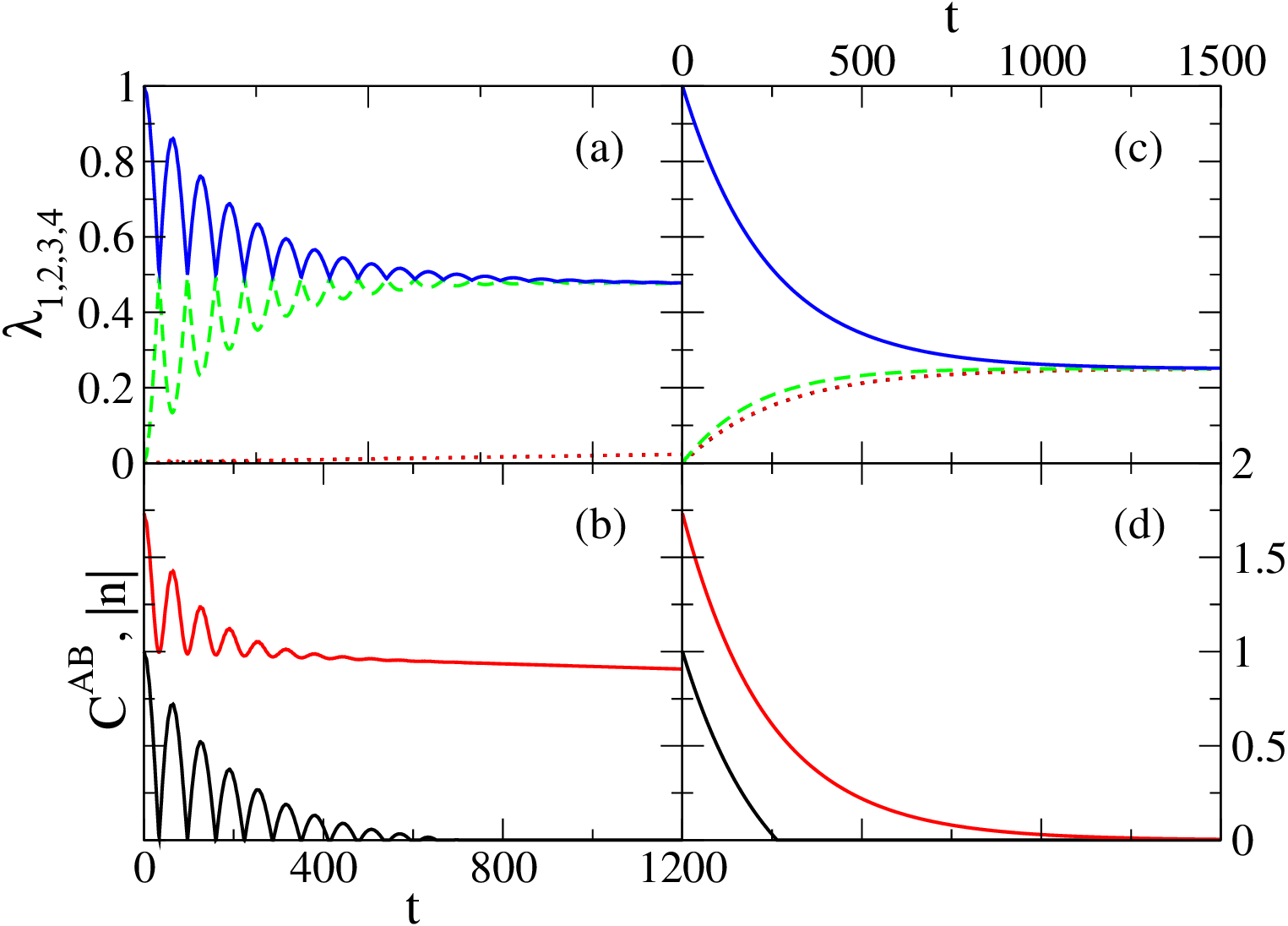}
\caption{Mixed noise for the two-one model.  Top panels (a) and (c): square
roots of eigenvalues of $\protect\rho \tilde{\protect\rho}(t)$.  Bottom
panels (b) and (d): magnitude of Bloch vector $n(t)$ above and concurrence $%
C^{AB}(t)$ below.  The curves are\ calculated from Eqs. \ref{eq:pd_21}-\ref{eq:C_pure} and confirmed
by numerical simulations. Both qubits are operated at an \textbf{%
intermediate working point} and only one of them is connected to the RTN,
i.e. $g_{1}=0.1$, $g_{2}=0$, $\protect\theta =\protect\pi /3,$ $\protect\phi %
=\protect\pi /2$. The initial state is set to $\left\vert \Phi
^{+}\right\rangle $. In (a) and (b), $\protect\gamma =0.005$ and the RTN is
in the strong-coupling region while in (c) and (d), $\protect\gamma =0.5$
and the RTN is in the weak-coupling region. $\protect\lambda _{1}$ is
plotted as a solid blue line, $\protect\lambda _{2}$ as a dashed green line; 
$\protect\lambda _{3}$ as a dottd red line, and $\protect\lambda _{4}$ as a
dotted black line. $n(t)$ is plotted as a solid red line and $C^{AB}(t)$ as
a solid black line.  Time is in the unit $1/B_{0}$. }
\label{fig:aw_Bell}
\end{figure}

For $\theta >0,$ \textit{intermediate working points}, $\Gamma _{1}>0$ and $
\xi \left( t\rightarrow \infty \right) =1/2$.  $\lambda _{3}=\lambda _{4}$
increase monotonically with time.  In fact all eigenvalues approach the
same limit: $\lambda _{1,2,3,4}\left( t\rightarrow \infty \right) =\alpha 
\sqrt{1-\alpha ^{2}}/2$  as seen in Fig. \ref{fig:aw_Bell} (a,c).  If $
\theta >0,$ then ESD is inevitable. All $\lambda $'s have $\alpha \sqrt{
1-\alpha ^{2}}/2$ as their long time limit. At strong coupling, there are a
finite number of revivals, while at weak coupling, ESD occurs at a finite
time and no revival occurs.  This generic behavior is the same at all intermediate working points.

As we begin in a pure state, we have $\left\vert \vec{n}\left( t=0\right)
\right\vert =\sqrt{3}$.  At long times the magnitude of the Bloch vector
approaches a finite limit in the dephasing case because $n_{3}=\left\langle
\sigma _{0}\otimes \sigma _{3}\right\rangle $, proportional to the
expectation value of the $z$-component of the spin for qubit $B$, is
independent of time.  This is due to the cylindrical symmetry of the
Hamiltonian: $\left[ H,\sigma _{0}\otimes \sigma _{3}\right] =0$.  In the
case $\theta >0$, $\left\vert \vec{n}\right\vert $ decays to zero, though
this happens on a time scale longer than is shown in Fig. \ref{fig:pd_Bell}.
 Nevertheless, in both cases the time dependence of $\left\vert \vec{n}
\left( t\right) \right\vert $ tracks the time dependence of $C^{AB}$ to a
large extent.  In particular, the oscillations in $C^{AB}\left( t\right) $
observed at strong coupling are also present in $\left\vert \vec{n}\left(
t\right) \right\vert $.  These oscillations might naively be supposed to
come from oscillations between entangled and product states; the fact that
the oscillations are also present in $\left\vert \vec{n}\right\vert $ means
that the disentanglement is coming essentially from mixing, even though it
is non-monotonic. 

Notice that $\left\vert \vec{n}\right\vert ,$ unlike $C^{AB},$ is a
continuous function of the elements of the density matrix $\rho $; as a
result it does not suffer sudden death but rather decays exponentially. 
The oscillations in $C^{AB}$ are not as long-lived as those in $\left\vert 
\vec{n}\right\vert $.  This is due to the fact that once the envelope of $
\zeta $ is less than $\xi ,$ ESD kills off $C^{AB}$.  No such effect occurs
for $\left\vert \vec{n}\right\vert $. 

\subsection{Mixed States}

We have found that for pure initial states, the disentanglement comes mainly
from mixing.  Additional insight is gained by taking mixed initial states.
 We will use the "extended Werner states" \cite{Werner_1989,Bellomo_2008}
as initial states, 
\begin{align}
w_{r}^{\Phi }(0)& =r~\left\vert \Phi \right\rangle \left\langle \Phi
\right\vert +\frac{1-r}{4}I_{4}, \\
w_{r}^{\Psi }(0)& =r~\left\vert \Psi \right\rangle \left\langle \Psi
\right\vert +\frac{1-r}{4}I_{4},
\end{align}%
where $0<r<1$ measures the purity of the initial states. $r=1$ gives the
(pure) generalized Bell states of the previous section and $r=0$ is the
fully mixed state $\rho =I_{4}/4.$ \ $\left\vert \vec{n}\left( t=0\right)
\right\vert =\sqrt{3}r.$

The generalized Bloch vectors for the Werner states $\left( \vec{n}_{w_r^{\Phi
}},\vec{n}_{w_r^{\Psi }}\right) $ are related to the those for the generalized
Bell states $\left( \vec{n}_{|\Phi >},\vec{n}_{|\Psi >}\right) $ by simple
scaling 
\begin{align}
\vec{n}_{w^{\Phi }_r}(t)& =r~\vec{n}_{|\Phi >}(t),  \label{eq:scaling} \\
\vec{n}_{w^{\Psi }_r}(t)& =r~\vec{n}_{|\Psi >}(t).
\end{align}

For both initial states $w_{r}^{\Phi }(0)$ and $w_{r}^{\Psi }(0)$, the $%
\lambda $'s are given by 
\begin{align}
{\lambda _{1}}=& r\alpha \Vert \beta \Vert \left( \tilde{\xi}+|\zeta
|\right) , \\
{\lambda _{2}}=& r\alpha \Vert \beta \Vert \left( \tilde{\xi}-|\zeta
|\right) , \\
{\lambda _{3}}=& {\lambda _{4}}=r\alpha \Vert \beta \Vert ~\xi
\end{align}%
where 
\begin{equation*}
\tilde{\xi}=\frac{\sqrt{(1+re^{-\Gamma _{1}t})^{2}-(1+e^{-\Gamma
_{1}t})^{2}r^{2}(2\alpha ^{2}-1)^{2}}}{4r\alpha \sqrt{1-\alpha ^{2}}},
\end{equation*}%
and the relaxation function is given by 
\begin{equation*}
\xi =\frac{\sqrt{(1-re^{-\Gamma _{1}t})^{2}-(1-e^{-\Gamma
_{1}t})^{2}r^{2}(2\alpha ^{2}-1)^{2}}}{4r\alpha \sqrt{1-\alpha ^{2}}},
\end{equation*}%
which depends both on the relaxation rate $\Gamma _{1}$ and the initial
conditions of the qubit.

The concurrence is then 
\begin{equation}
C^{AB}=\max \left\{ 0,~2r\alpha \sqrt{1-\alpha ^{2}}\left( |\zeta |-\xi
\right) \right\} ,  \label{eq:C_aw_21}
\end{equation}

If $\Gamma _{1}\neq 0$, i.e., if the qubit is operated at intermediate
working point, the long time limit of $\xi $ is 
\begin{equation}
\xi (t\rightarrow \infty )=\frac{\sqrt{1-r^{2}(2\alpha ^{2}-1)^{2}}}{%
4r\alpha \sqrt{1-\alpha ^{2}}}.
\end{equation}
If $\Gamma _{1}=0$ (pure dephasing), 
\begin{align}
\xi (t)=\frac{{1-r}}{4r\alpha \sqrt{1-\alpha ^{2}}}.\label{eq:pd_xi}
\end{align}
Thus the only situation where ESD does not happen is when $r=1$ and $\Gamma
_{1}=0$.  ESD is thus essentially a universal behavior if qubits are
subjected to RTNs.

\begin{figure}[tb]
\centering\includegraphics*[width=\linewidth]{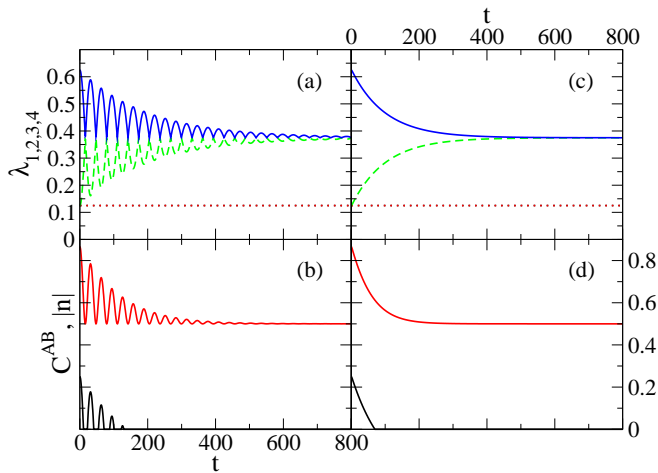}
\caption{Top panels (a) and (c) square roots of eigenvalues of $\protect
\rho \tilde{\protect\rho}(t)$.  Bottom panels (b)\ and (d): Bloch vector $%
n(t)$ above and concurrence $C^{AB}(t)$ below.  The curves are calculated
from Eqs. \ref{eq:scaling}-\ref{eq:C_aw_21} and confirmed by numerical simulations. Both qubits are
operated at \textbf{pure dephasing point} and only one of them is connected
to the RTN, i.e. $g_{1}=0.1$, $g_{2}=0$, $\protect\theta =0$. The initial
state is set to $\protect\rho _{r}^{\Phi ^{+}}$ with $r=0.5$. In (a) and
(b), $\protect\gamma =0.005$ and the RTN is in the strong-coupling region
while in (c) and (d), $\protect\gamma =0.5$ and the RTN is in the
weak-coupling region. $\protect\lambda _{1}$ is plotted as solid blue line, $%
\protect\lambda _{2}$ dashed green line, $\protect\lambda _{3}$ dotted red
line, $\protect\lambda _{4}$ dotted black line, $n(t)$ solid red line and $%
C^{AB}(t)$ solid black line. Note $\protect\lambda _{3}=\protect\lambda %
_{4}\neq 0$ and ESD occurs in both cases. Time is in the unit $1/B_{0}$.}
\label{fig:pd_Werner}
\end{figure}

\begin{figure}[tb]
\centering\includegraphics*[width=\linewidth]{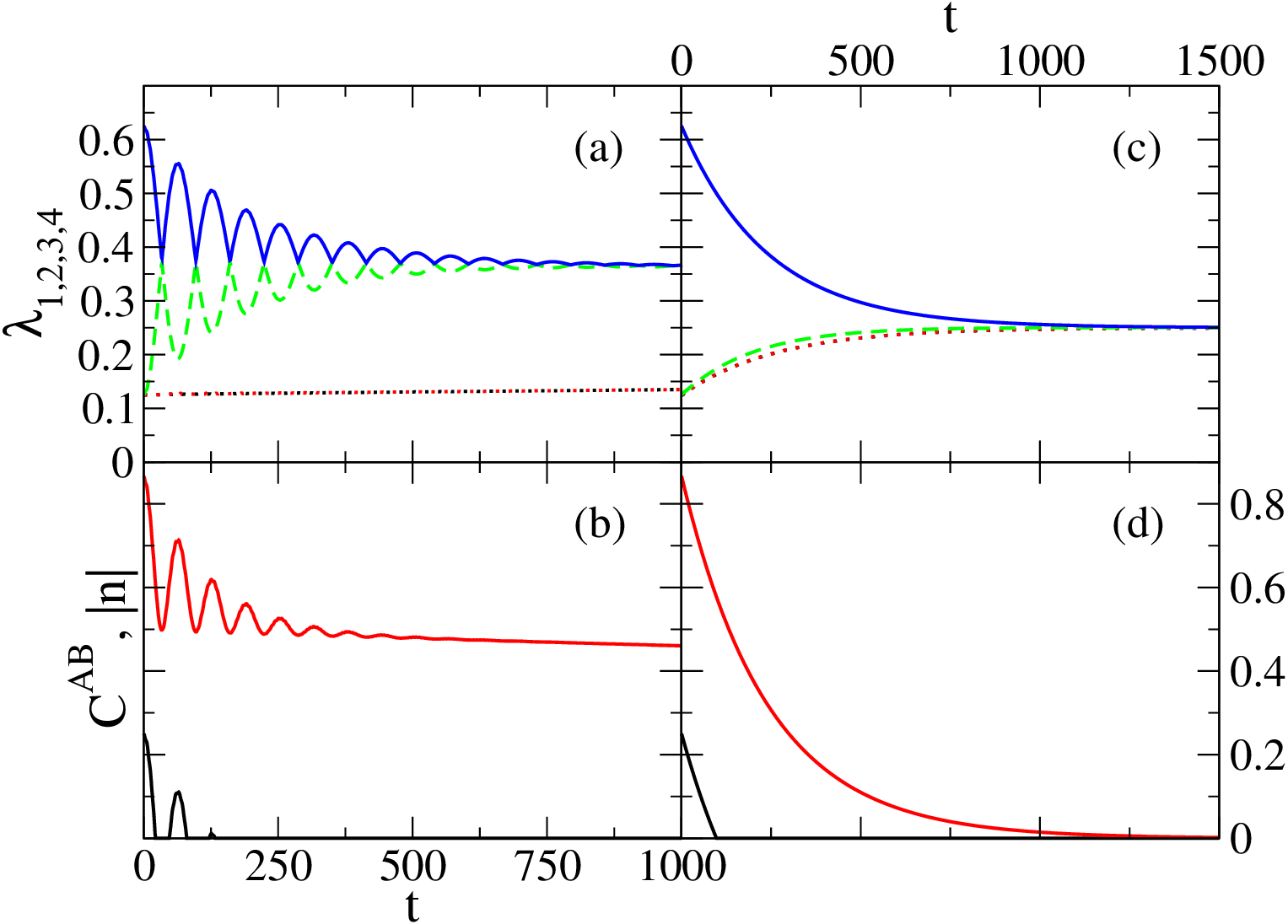}
\caption{Top panels (a) and (c): square roots of eigenvalues of $\protect%
\rho \tilde{\protect\rho}(t)$.  Bottom panels (b) and (d): magnitude of
Bloch vector $n(t)$ above and and concurrence $C^{AB}(t)$ below.  The
curves are calcuated from Eqs, \ref{eq:scaling}-\ref{eq:C_aw_21} and confirmed by numerical simulations.
Both qubits are operated at \textbf{intermediate working point} and only one
of them is connected to the RTN, i.e. $g_{1}=0.1,g_{2}=0$, $\protect\theta =%
\protect\pi /3$, $\protect\phi =\protect\pi /2$. The initial state is set to 
$\protect\rho _{r}^{\Phi ^{+}}$ with $r=0.5$. In (a) and (b), $\protect%
\gamma =0.005$ and the RTN is in the strong-coupling region while in (c) and
(d), $\protect\gamma =0.5$ and the RTN is in the weak-coupling region. $%
\protect\lambda _{1}$ is plotted as solid blue line, $\protect\lambda _{2}$
dashed green line, $\protect\lambda _{3}$ dotted red line, $\protect\lambda %
_{4}$ dotted black line, $n(t)$ solid red line and $C^{AB}(t)$ solid black
line. Note $\protect\lambda _{3}=\protect\lambda _{4}\neq 0$ and ESD occurs
in both cases. Time is in the unit $1/B_{0}$.}
\label{fig:aw_Werner}
\end{figure}

In Fig. \ref{fig:pd_Werner} and \ref{fig:aw_Werner}, we can see in greater
detail how the introduction of $r,$ i.e. the interpolation with $I_{4}$ in
the initial density matrix, changes the situation.  First, it lifts $%
\lambda _{3}$ and $\lambda _{4}$ to finite values.  This causes ESD to
happen even at the pure dephasing point.  $r$ is essentially a radial
variable in $\vec{n}$-space.  Comparing Fig. \ref{fig:pd_Bell} to Fig. \ref{fig:pd_Werner} (pure dephasing
noise), we find, in agreement with Eq. \ref{eq:scaling} that $\vec{n}$ is
simply proportional to $r$.  Thus the decay time of $\left\vert \vec{n}%
\right\vert $ is unchanged.  The dependence of $C^{AB}$ on $r,$ however, is
more complicated.  There is an overall proportionality, but the relaxation
function $\xi $ also dpends on $r$.  This means that the decay time of $%
C^{AB}$ is reduced when $r=0.5$ as compared to when $r=1$ (pure initial
state).  This is in agreement with Eq. \ref{eq:C_aw_21}.  Comparison of
Figs. \ref{fig:aw_Bell} and \ref{fig:aw_Werner} (arbitrary working point) shows that this qualitative
behavior does not change when we have mixed noise.

This says something important about the geometry of the 15-dimensional space
in which $\vec{n}$ lives.  If we move radially in this space from a point
with $\left\vert \vec{n}\right\vert =\sqrt{3}$ and $C^{AB}=1$ (a pure state
with maximal entanglement) to the origin (the maximally mixed state) along a
path determined by our model the entanglement diminishes monotonically, but
not smoothly, to $0$.  If we move on the surface of the sphere of pure states
(changing angular variables only), it is obviously possible to move
continuously from an entangled pure state with $C^{AB}=1$ to a product
(separable) state that would have $C^{AB}=0$.  We may summarize this by
saying that in our model purity is a radial variable in $\vec{n}$-space
while entanglement is, roughly speaking, the product of the purity and an
angular variable. 

\section{two-two model}

\label{sec:2_2_model} In this section, both qubits are subject to RTNs and
the two RTNs are not correlated and do not necessarily have the same
prameters $g$ and $\gamma $. The transfer matrix for this model is 
\begin{equation*}
\underline{T}(t)=\underline{R}_{\text{RTN}}^{A}(t)\otimes \underline{R}_{\text{RTN}}^{B}(t).
\end{equation*}

We only consider the extended Werner states here since the generalized Bell
states are included as special cases.

With $w_{r}^{\Phi }$ as initial state, the non-zero components of the
generalized Bloch vector are 
\begin{align}
n_{3}(t)=& r~(2\alpha ^{2}-1)~e^{-\Gamma _{1}^{B}t}, \label{eq:n_22}\\
n_{5}(t)=& 2r~\zeta ^{A}\zeta ^{B}\alpha 
\left[ \cos (2B_{0}t)\text{Re}~\beta +\sin (2B_{0}t)\text{Im}~\beta \right]
\\
n_{6}(t)=& 2r~\zeta ^{A}\zeta ^{B}\alpha 
\left[ \cos (2B_{0}t)\text{Im}~\beta -\sin (2B_{0}t)\text{Re}~\beta \right]
\\
n_{9}(t)=& n_{6}(t), \\
n_{10}(t)=& -n_{5}(t), \\
n_{12}(t)=& r~(2\alpha ^{2}-1)~e^{-\Gamma _{1}^{A}t}, \\
n_{15}(t)=& r~e^{-(\Gamma _{1}^{A}+\Gamma _{1}^{B})t}.
\end{align}

With $w_{r}^{\Psi }$ as initial state, the non-zero components of the
generalized Bloch vector are 
\begin{align}
n_{3}(t)=& r (1-2\alpha ^{2})~e^{-\Gamma _{1}^{B}t}, \\
n_{5}(t)=& 2r\alpha~\text{Re}~\beta ~\zeta ^{A}\zeta ^{B}, \\
n_{6}(t)=& -2r\alpha~\text{Im}~\beta~\zeta ^{A}\zeta ^{B},\\
n_{9}(t)=& -n_{6}(t), \\
n_{10}(t)=& n_{5}(t), \\
n_{12}(t)=& r (2\alpha ^{2}-1)~e^{-\Gamma _{1}^{A}t}, \\
n_{15}(t)=& -r e^{-(\Gamma _{1}^{A}+\Gamma _{1}^{B})t}.
\end{align}
Note there is no $B_0$ dependence in this case.

The square roots of eigenvalues of $\rho_{AB}\tilde\rho_{AB}$ for both initial states are 
\begin{align}
{\lambda_1} &= r\alpha\|\beta\|\left(\tilde{\xi}+|\zeta^A\zeta^B|\right) \\
{\lambda_2} &=r\alpha\|\beta\|\left(\tilde{\xi}-|\zeta^A\zeta^B|\right) \\
{\lambda_3}&= {\lambda_4}= r\alpha\|\beta\|~\xi
\end{align}
where 
\begin{align}
\tilde{\xi}= \frac{\sqrt{\left(1+re^{-(\Gamma_1^A+\Gamma_1^B)t}%
\right)^2-r^2(2\alpha^2-1)^2\left(e^{-\Gamma_1^At}+e^{-\Gamma_1^Bt}\right)^2}%
}{4r\alpha\sqrt{1-\alpha^2}}
\end{align}
and the relaxation function is 
\begin{align}
\xi= \frac{\sqrt{\left(1-re^{-(\Gamma_1^A+\Gamma_1^B)t}\right)^2-r^2(2%
\alpha^2-1)^2\left(e^{-\Gamma_1^At}-e^{-\Gamma_1^Bt}\right)^2}}{4r\alpha%
\sqrt{1-\alpha^2}}.
\end{align}

The concurrence is given by 
\begin{equation}
C^{AB}=\max \left\{ 0,2r\alpha \sqrt{1-\alpha ^{2}}\left[ \left\vert \zeta
^{AB}\left( t\right) \right\vert -\xi \left( t\right) \right] \right\}\label{eq:C_22}
\end{equation}
where 
\begin{equation*}
\zeta ^{AB}\left( t\right) =\zeta ^{A}\left( t\right) \zeta ^{B}\left(
t\right) .
\end{equation*}%
This simple product form is due to the independence between the two qubits.

\begin{figure}[tb]
\centering\includegraphics*[width=\linewidth]{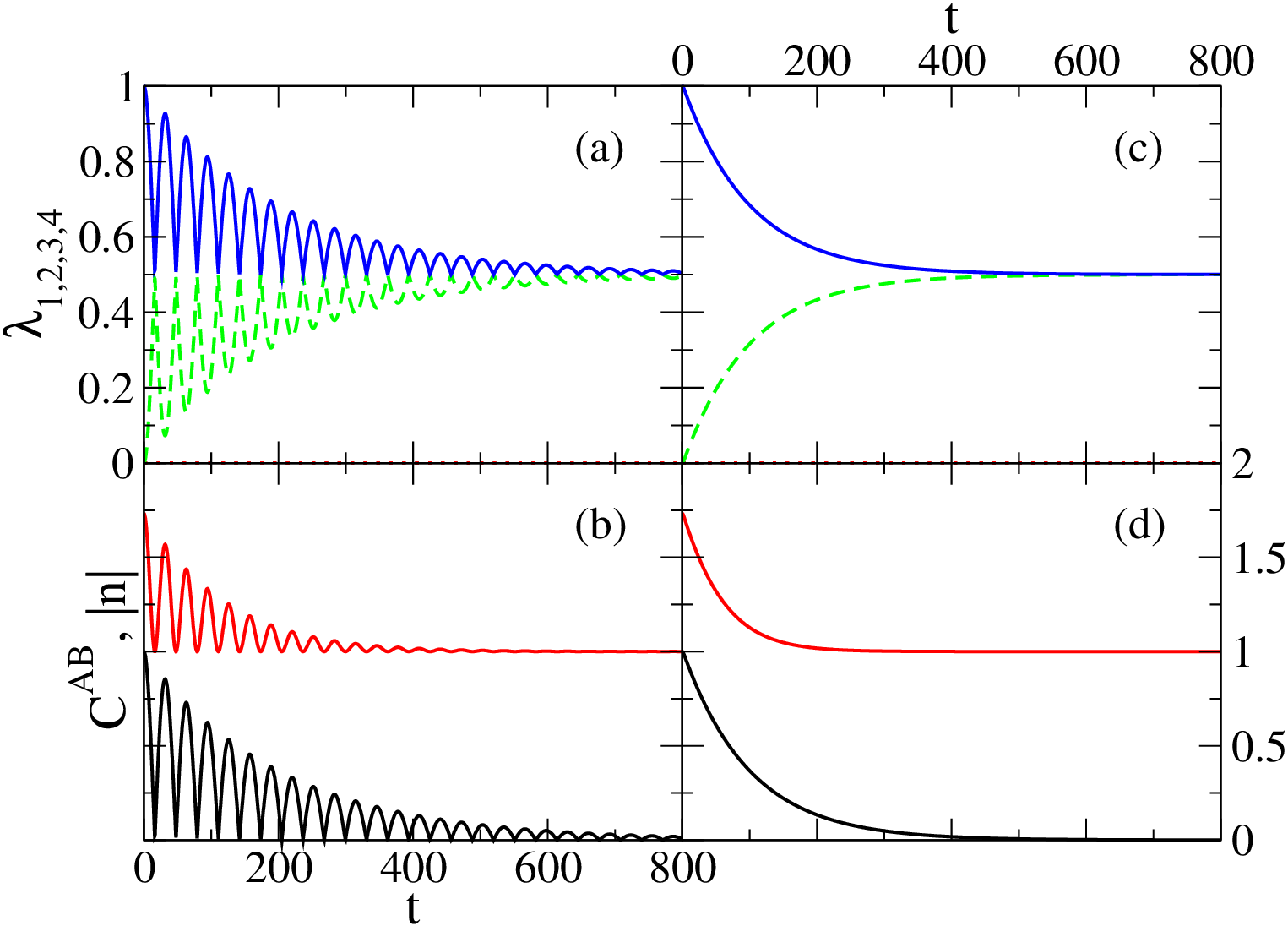}
\caption{Top panels (a) and (c): square roots of eigenvalues of $\protect
\rho \tilde{\protect\rho}(t)$.  Bottom panels (b) and (d): magnitude of
Bloch vector $n(t)$ above and and concurrence $C^{AB}(t)$ below.  The
curves are calcuated from Eqs. \ref{eq:n_22}-\ref{eq:C_22} and confirmed by numerical simulations.
Both qubits are operated at \textbf{pure dephasing point} and both of them
are connected to the RTN, i.e. $g_{1}=0.1$, $g_{2}=0.1$, $\protect\theta =0$
. The initial state is set to $\left\vert \Phi ^{+}\right\rangle $. In (a)
and (b), $\protect\gamma =0.005$ and the RTN is in the strong-coupling
region while in (c) and (d), $\protect\gamma =0.5$ and the RTN is in the
weak-coupling region. $\protect\lambda _{1}$ is plotted as solid blue line, $
\protect\lambda _{2}$ dashed green line, $\protect\lambda _{3}$ dotted red
line, $\protect\lambda _{4}$ dotted black line, $n(t)$ solid red line and $
C^{AB}(t)$ solid black line. Note $\protect\lambda _{3}=\protect\lambda %
_{4}=0$ and ESD does not occur in both cases. Time is in the unit $1/B_{0}$.}
\label{fig:2_2_pd_Bell}
\end{figure}

The long time limit of the relaxation function is 
\begin{equation*}
\xi (t\rightarrow \infty )=\frac{1}{4r\alpha \sqrt{1-\alpha ^{2}}}\quad 
\text{if }\Gamma _{1}^{A}\neq 0, \Gamma _{1}^{B}\neq 0.
\end{equation*}%
If $\Gamma _{1}^{A}=\Gamma _{1}^{B}=0$, Eq.\ref{eq:pd_xi} is recovered and $%
r=1$ prevents ESD from happening, as seen in Fig. \ref{fig:2_2_pd_Bell}. 

The qualitative behavior of all quantities is rather similar in Figs. \ref{fig:pd_Bell}, \ref{fig:pd_Werner} and \ref{fig:2_2_pd_Bell},
all referring to dephasing noise.  This confirms a qualitative
picture in which both qubits undergo a random walk in their respective
Hilbert spaces; relative variables therefore also undergo a random
walk, but faster.  Quantities like $C^{AB}$ that depend on the relative variables have
a faster decay time.  (Note the difference in the time scales on 
Figs. \ref{fig:pd_Bell} and \ref{fig:2_2_pd_Bell}.) 

Finally we note the norm of the Bloch vector $|n|$ resembles concurrence in all cases considered. 
This can be seen from  the explicit expression
\begin{align}
|n| = r\sqrt{\begin{array}{c}
    8\alpha^2(1-\alpha^2)\left(\zeta^{AB}\right)^2
    + e^{-2\left( \Gamma_1^At+\Gamma_1^Bt \right)}\\
+(1-2\alpha^2)^2\left[e^{-2\Gamma_1^At}+e^{-2\Gamma_1^Bt}\right]
\end{array}}.
\end{align}
Since the generalized Bloch vector fully describes the system, a geometric picture of $\vec n$ for entanglement might be possible.
To our knowledge, such a description is not yet available except for some special parameterized states \cite{Bertlmann} or close to $\vec{n}=\vec{0}$ \cite{mixed}.

\section{discussion and conclusion}

\label{sec:phase}

\begin{figure}[tb]
\centering\includegraphics*[width=.8\linewidth]{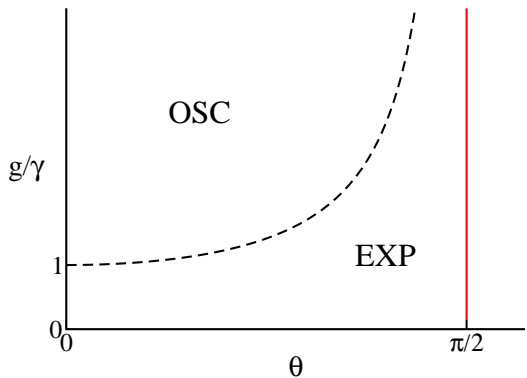}
\caption{"Phase diagram" of the behavior of $C^{AB}(t)$, given 'extended'
Werner state as initial state. In the upper region, we have ESD and revival
before $C^{AB}$ goes permanently to zero.  In the lower region, $C^{AB}$
dies just once and for all. } \label{fig:phase_d}
\end{figure}

The most important conclusion of the paper is that the disentangling effect
of non-Markovian, or strongly-coupled, noise and the effect of Markovian, or
weakly-coupled, noise is qualitatively different.  In Sec. \ref%
{sec:2_1_model} and \ref{sec:2_2_model}, we see $C^{AB}(t)$ can take on two
forms before ESD occurs: oscillatory and exponential, depending on the
coupling of the RTN.  By numerical exploration, we have constructed the
"phase diagram" in Fig. \ref{fig:phase_d}, where the boundary is given by 
\begin{equation*}
g/\gamma =\sec \theta .
\end{equation*}

The oscillatory behavior arises from looping of $\vec{n}$ on the Bloch
sphere.  This can only occur if the noise is slow enough that the topology
if the sphere is fully explored before the Bloch vector decays entirely. 
If the noise is fast, the relaxation moves $\vec{n}$ along a radial path to
the origin no looping occurs.

If the noise acts only on one qubit, the two-one model, the situation can be
analyzed in some detail.  Qubit $B$ is not subject to RTN and is stationary
in the rotating frame. It effectively serves as a reference and the
two-qubit concurrence is fully determined by qubit $A$, as seen in Eq. \ref{eq:C_pure}.

In the single-qubit Bloch sphere picture, 
\begin{align}
|\zeta |& =\rho _{\text{A}}|\sin \theta _{\text{A}}| \\
e^{-\Gamma _{1}t}& =\rho _{\text{A}}\cos \theta _{\text{A}},
\end{align}%
where $\rho _{\text{A}}$ and $\theta _{\text{A}}$ are the length and polar
angle of the three dimensional Bloch vector of qubit $A$.

\begin{figure}[tb]
\centering\includegraphics*[width=\linewidth]{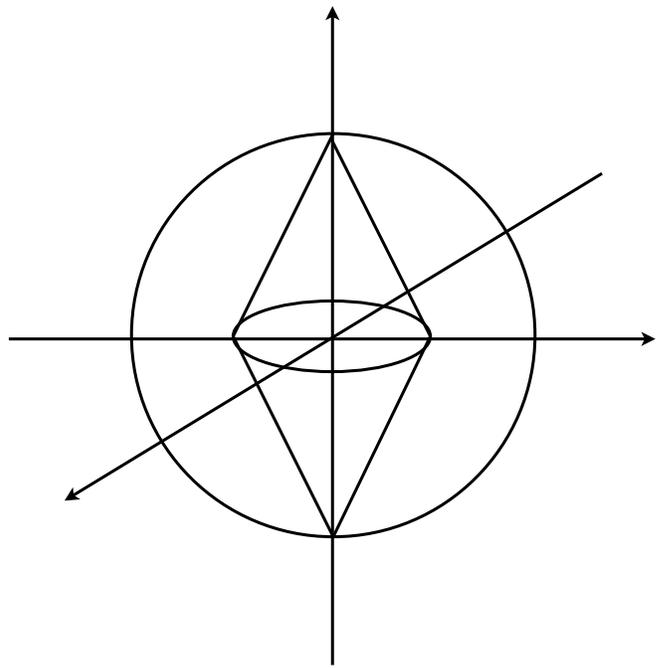}
\caption{Single-qubit Bloch representation of the two-one model. Initial
states are the generalized Bell states. Inside the cone C$^{AB}=0$.}
\label{fig:pd_bloch}
\end{figure}

Given generalized Bell states as initial state, $C^{AB}=0$ is equivalent to 
\begin{equation*}
2\rho _{\text{A}}|\sin \theta _{\text{A}}|+\rho _{\text{A}}\cos \theta _{%
\text{A}}\leq 1.
\end{equation*}%
Geometrically, it means $C^{AB}=0$ as long as qubit $A$'s Bloch vector falls
inside the cone shown in Fig. \ref{fig:pd_bloch}.

In the two-two model, however, both qubits have nontrivial time evolution.
This simple one-qubit picture for concurrence then does not work and one
needs to treat the full 15-dimensional Bloch vector for the whole system.

One important question is the relation of this work to previous results on single qubits \cite{Zhou_2009,Galperin,Paladino}. 
The oscillations that occur in $|\vec n|$ and $C^{AB}$ are clearly related to the noise-induced looping on the single-qubit Bloch sphere. 
For example, they have the same period.
However, in the single-qubit case these oscillations occur in the tails of an overall Gaussian decay. 
They are much more pronounced in $|\vec n|$ and $C^{AB}$.

Entanglement revival was introduced in Ref. \cite{Ficek_2006} and later on shown to exist in different systems \cite{Bellomo_2007,Maniscalco}.
It has sometimes been attributed to back-action from the non-Markovian environment \cite{Mazzola}.
There is no back-action in our model so this cannot be a general statement.
Furthermore, the revival seen in the present work is a simple oscillation and does not arise from any constructive interference of multiple reservoirs.

In conclusion, we used the quasi-Hamiltonian method to study the entanglement dynamics of two non-interacting qubits subject to uncorrelated RTNs, utilizing the generalized
(15-dimensional) Bloch vector $\vec{n}$.  This turns out to be
very well suited to determining entanglement measures such as the
concurrence, since $\vec{n}$ has a rather direct relation to entanglement. 
We found in our work that disentanglement caused by classical noise on
2-qubit systems falls into two distinct categories.  In the Markovian noise case
familiar from perturbation (Redfield) theory, the motion of $\vec{n}$ is
essentially radial and ESD happens except in special cases.  The time scale
of ESD is similar to the time scale of exponential decay of $\left\vert \vec{
n}\right\vert$.  In the Markovian noise case there is a combination of radial
and angular motion of $\vec{n}$; $\left\vert \vec{n}\right\vert $ typically
shows oscillatory behavior, while the concurrence undergoes ESD \textit{and}
revival. {The quasi-Hamiltonian method provides us a flexible way to deal with independent qubits and uncorrelated noises. 
The formalism for multiple qubits has been established in Ref. \cite{Joynt_2009} and it is therefore straightforward to extend the present work to this case.
Other future work would be to explore the effects of inter-qubit coupling and noise correlations.}

\begin{acknowledgments}
This work was supported by the National Science Foundation, Grant Nos.
NSF-ECS-0524253 and NSF-FRG-0805045, by the Defense Advanced Research
Projects Agency QuEST program, and by ARO and LPS Grant No. W911NF-08-1-0482. A. Lang was supported by a Hilldale Research Fellowship and the QuEST grant number QuEST grant \#MSN118850.
\end{acknowledgments}


\begin{thebibliography}{99}
\bibitem{Horodecki_2009} R. Horodecki, P. Horodecki, M. Horodecki, and K.
Horodecki, Rev. Mod. Phys. \textbf{81}, 865 (2009).

\bibitem{crypto} N. Gisin, G. Ribordy, W. Tittel, and H. Zbinden, Rev. Mod.
Phys. \textbf{74}, 145 (2002).

\bibitem{Bennett_1992} C.H. Bennett and S.J. Wiesner, Phys. Rev. Lett. 
\textbf{69}, 2881 (1992).

\bibitem{Bennett_1993} C.H. Bennett, G. Brassard, C. Crépeau, R. Jozsa, A.
Peres, and W.K. Wooters, Phys. Rev. Lett. \textbf{70}, 1895 (1993).

\bibitem{algo} P.W. Shor, SIAM J. Comp. \textbf{26}, 1484 (1997); L.K.
Grover, Phys. Rev. Lett. \textbf{79}, 325 (1997).

\bibitem{Yu_2009_review} T. Yu and J.H. Eberly, Science \textbf{323}, 598
(2009).

\bibitem{ESD_exp} M.P. Almeida, F. de Melo, M. Hor-Meyll, A. Salles, S.P.
Walborn, P.H. Souto Ribeiro, and L. Davidovich, Science \textbf{316}, 579
(2007); J. Laurat, K.S. Choi, H. Deng, C.W. Chou, H.J. Kimble, Phys. Rev.
Lett. \textbf{99}, 180504 (2007).

\bibitem{Yu_Eberly} T. Yu, J.H. Eberly, Phys. Rev. Lett. \textbf{93}, 140404 (2004);
 \textit{ibid}. \textbf{97}, 140403 (2006).

\bibitem{Yu_2006} T. Yu, J.H. Eberly, Opt. Commun. \textbf{264}, 393 (2006).

\bibitem{Santos_2006} M.F. Santos, P. Milman, L. Davidovich, and N. Zagury,
Phys. Rev. A \textbf{73}, 040305(R) (2006).

\bibitem{Ficek_2006} Z. Ficek and R. Tana\'s, Phys. Rev. A \textbf{74},
024304 (2006).

\bibitem{Kogan} S. Kogan, \textit{Electronic Noise and Fluctuations in
Solids}, (Cambridge Univ. Press, Cambridge, 1996).

\bibitem{Weissman_1988} M.B. Weissman Rev. Mod. Phys. \textbf{60}, 537
(1988); P. Dutta and P.M. Horn, Rev. Mod. Phys. \textbf{53}, 497 (1981).

\bibitem{Yoshihara_2006} F. Yoshihara, K. Harrabi, A.O. Niskanen, Y.
Nakamura, and J.S. Tsai, Phys. Rev. Lett. \textbf{97}, 167001 (2006).

\bibitem{Kakuyanagi_2007} K. Kakuyanagi, T. Meno, S. Saito, H. Nakano, K.
Semba, H. Takayanagi, F. Deppe, and A. Shnirman, Phys. Rev. Lett. \textbf{98}%
, 047004 (2007).

\bibitem{Nakamura_2002} Y. Nakamura, Yu.A. Pashkin, T. Yamamoto, and J.S.
Tsai, Phys. Rev. Lett. \textbf{88}, 047901 (2002).

\bibitem{1/f} F. C. Wellstood, C. Urbina, and J. Clarke, Appl. Phys. Lett. 
\textbf{99}, 187006 (2007); R. C. Bialczak, R. McDermott, M. Ansmann, M.
Hofheinz, N. Katz, E. Lucero, M. Neeley, A.D. O'Connell, H. Wang, A.N.
Cleland, and J.M. Martinis, Phys. Rev. Lett. \textbf{99}, 187006 (2007); D.
J. Van Harlingen, T.L. Robertson, B.L.T. Plourde, P.A. Reichardt, T.A.
Crane, and J. Clarke, Phys. Rev. B \textbf{70}, 064517 (2004).

\bibitem{Khaetskii_2002} A.V. Khaetskii, D. Loss, and L. Glazman, Phys. Rev.
Lett. \textbf{88}, 186802 (2002).

\bibitem{Bellomo_2007} B. Bellomo, R. Lo Franco, and G. Compagno, Phys. Rev.
Lett. \textbf{99}, 160502 (2007);

\bibitem{Bellomo_2008} B. Bellomo, R. Lo Franco, and G. Compagno, Phys. Rev.
A \textbf{77}, 032342 (2008).

\bibitem{Dajka_2008} J. Dajka, M. Mierzejewski, and J. Luczka, Phys. Rev. A 
\textbf{77}, 042316 (2008).

\bibitem{Mazzola} L. Mazzola, S. Maniscalco, J. Piilo, K.-A. Suominen, and B. M. Garraway, Phys. Rev. A \textbf{79}, 042302 (2009).

\bibitem{Yu_2009} T. Yu and J.H. Eberly, Opt. Commun. \textbf{283}, 676 (2010).

\bibitem{Hollenberg} M.J. Testolin, J.H. Cole, and L.C.L. Hollenberg, Phys.
Rev. A \textbf{40}, 042326 (2009).

\bibitem{Yu_2002} T. Yu and J.H. Eberly, Phys. Rev. B \textbf{66}, 193306
(2002).

\bibitem{Roszak_2006} K. Roszak and P. Machnikowski, Phys. Rev. A \textbf{73}%
, 022313 (2006).

\bibitem{Ann_2007} K. Ann and G. Jaeger, Phys. Rev. B \textbf{75}, 115307
(2007).

\bibitem{Cao_2008} X. Cao and H. Zheng, Phys. Rev. A \textbf{77}, 022320 
(2008).

\bibitem{Cheng_2008} B. Cheng, Q.H. Wang, and R. Joynt, Phys. Rev. A \textbf{%
78}, 022313(2008).

\bibitem{Joynt_2009} R. Joynt, D. Zhou, and Q.H. Wang, arXiv:0906.2843.

\bibitem{Zhou_2009} D. Zhou and R. Joynt, Phys. Rev. A \textbf{81}, 010103(R) (2010).

\bibitem{Galperin} Y.M. Galperin, B.L. Altshuler, J. Bergli, and D.V.
Shantsev, Phys. Rev. Lett. \textbf{96}, 097009 (2006); Y.M. Galperin, B.L.
Altshuler, J. Bergli, D. Shantsev and V. Vinokur, Phys. Rev. B \textbf{76},
064531 (2007).

\bibitem{Paladino} E. Paladino, L. Faoro, G. Falci, and R. Fazio, Phys. Rev.
Lett. \textbf{88}, 228304 (2002); G. Falci, A. D'Arrigo, A. Mastellone, and
E. Paladino, Phys. Rev. Lett. \textbf{94}, 167002 (2005).

\bibitem{Byrd} M.S. Byrd and N. Khaneja, Phys. Rev. A \textbf{68}, 062322
(2003); note our normalization of the Bloch vector differs from this
reference.

\bibitem{Braun} D. Braun, Phys. Rev. Lett. \textbf{89}, 277901 (2002); M.S. Kim, J. Lee, D. Ahn, and P.L. Knight, Phys. Rev. A \textbf{65}, 040101(R) (2002); J.P. Paz and A.J. Roncaglia, Phys. Rev. Lett. \textbf{100}, 220401 (2008).

\bibitem{Slichter_1996} C. P. Slichter, \textit{Principles of Magnetic
Resonance}, 3rd ed. (Springer, New York, 1996).

\bibitem{Wootters_1998} W.K. Wootters, Phys. Rev. Lett. \textbf{80}, 2245
(1998).

%\bibitem{Yu_2007} T. Yu and J.H. Eberly, Quantum Inf. Comput. \textbf{7}, 459 (2007). 
%\bibitem{Yonac} M. Y\"{o}nac, T. Yu, J.H. Eberly, J. Phys. B \textbf{39}, S621 (2006); \textit{ibid}., \textbf{40}, S45 (2007).

\bibitem{Werner_1989} R.F. Werner, Phys. Rev. A \textbf{40}, 4277 (1989).
\bibitem{Bertlmann} R.A. Bertlmann and P. Krammer, arXiv:0706.1743 (2007).
\bibitem{mixed} S.L. Braunstein, C.M. Caves, R. Jozsa, N. Linden, S. Popescu, and R. Schack, Phys. Rev. Lett. \textbf{83}, 1054 (1999); W. D\"ur, J.I. Cirac, and R. Tarrach, \textit{i.b.i.d} \textbf{83}, 3562 (1999); L. Gurvits and H. Barnum, Phys. Rev. A \textbf{68}, 042312 (2003); L. Gurvits and H. Barnum, \textit{i.b.i.d} \textbf{72}, 032322 (2005).
\bibitem{Maniscalco} S. Maniscalco, F. Francica, R.L. Zaffino, N. Lo Gullo, and F. Plastina, Phys. Rev. Lett. \textbf{100}, 090503 (2008).
\end{thebibliography}
\end{document}